# Particle acceleration through the resonance of high magnetic field and high frequency electromagnetic wave

Hong Liu[1,4], X. T. He[2,3], S. G. Chen[2] and W.Y.Zhang[2]

[1]*Graduate School, China Academy of Engineering Physics, Beijing P. O. Box 2101, Beijing 100088, P.R.China*
[2]*Institute of Applied Physics and Computational Mathematics, Beijing P. O. Box 8009, Beijing 100088, P.R.China*
[3]*Department of Physics, Zhejiang University, Hangzhou 310027, China*
[4]*[4]Basic Department, Beijing Materials Institute, Beijing 101149, China*

We propose a new particle acceleration mechanism. Electron can be accelerated to relativistic energy within a few electromagnetic wave cycles through the mechanism which is named electromagnetic and magnetic field resonance acceleration (EMRA). We find that the electron acceleration depends not only on the electromagnetic wave intensity, but also on the ratio between electron Larmor frequency and electromagnetic wave frequency. As the ratio approaches to unity, a clear resonance peak is observed, corresponding to the EMRA. Near the resonance regime, the strong magnetic fields still affect the electron acceleration dramatically. We derive an approximate analytical solution of the relativistic electron energy in adiabatic limit, which provides a full understanding of this phenomenon. In typical parameters of pulsar magnetospheres, the mechanism allows particles to increase their energies through the resonance of high magnetic field and high frequency electromagnetic wave in each electromagnetic wave period. The energy spectra of the accelerated particles exhibit the synchrotron radiation behavior. These can help to understand the remaining emission of high energy electron from radio pulsar within supernova remnant. The other potential application of our theory in fast ignition scheme of inertial confinement fusion is also discussed.

Keywords: acceleration of particles-pulsars: general- radiation mechanisms: nonthermal

PACS number 98.70.Sa, 41.75.Lx, 97.60.Gb

## I. INTRODUCTION

In the past two decades strong magnetic field caused much interesting both in astrophysics[1] and in laser-matter interaction[2], e.g. many novel and complex physics involved in ultraintense laser-plasma interaction studies especially in inertial confinement fusion (ICF), including relativistic self-focusing[3], explosive channel formation[4] and self-generated huge magnetic field[5]. A typical problem is to investigate the response of electron in the presence of strong quasistatic magnetic fields. Experiments[6,7] and three-dimensional (3D) particle-in-cell (PIC) simulations[8] clearly demonstrate that the fact of strong currents of energetic $10-100$ $MeV$ electrons manifest themselves in a giant quasistatic magnetic field with up to $100$ $MG$ amplitude. Recently the laboratory astrophysics has a long development with the high intensity ($>10^{20}W/cm^{-2}$), high density ($\approx 1kg/cc$), high electric field ($\approx 300 GV/m$) and high pressure ($\approx 10 Gbar$) available in intense laser facilities. From the physical parameters of laser-plasma interaction experiments, we discover a quick and efficient electron acceleration related with intense electromagnetic wave and strong magnetic fields. If the similar parameters can be satisfied in pulsar environments, the efficient acceleration mechanism can help to understand the continue emission of electron from radio pulsars within supernova remnants.

The radio emission mechanism for pulsars is not adequately understood. Form Ruderman *et al.* "inner-gap model"[9], B. zhang, *et al.*[10,11], G.J.Qiao *et al.*[12] the curvature radiation (CR) mode and the inverse Compton scattering (ICS) mode (including resonant ICP gap mode and the thermal ICS gap mode), the low-frequency and high-frequency electromagnetic waves can be predicted. These theories mentioned the pulsars polar gaps, sparks, and coherent microwave radiation also. Whatever the details of the emission mechanism, the properties of the low-frequency and high-frequency waves in relativistic pair plasma in the pulsar magnetosphere are of central importance for understanding the underlying processes in the formation of the radio spectrum. In this paper we concentrate on high-frequency electromagnetic wave case.

From above, we restrict our attention to a typical problem which is to investigate the response of electron in ultraintense electromagnetic wave plasma system in the presence of strong magnetic field. Using test particle model, we solve relativistic Lorentz force equations theoretically and experimentally. In our simulation, the electromagnetic wave is a circular polarized (CP) Gaussian profile. The magnetic field is considered as an axial constant field. A fully relativistic single particle code is developed to investigate the dynamical properties of the energetic electrons. We find that a rest electron can be accelerated to relativistic energy within a few electromagnetic wave cycle through a mechanism which is named electromagnetic and magnetic field resonance acceleration (EMRA). The electron acceleration depends not only on the electromagnetic wave intensity, but also on the ratio between electron Larmor frequency and electromagnetic wave frequency. As the ratio approaches to unity, a clear resonance peak is observed, corresponding to the EMRA. Near the resonance regime, the strong magnetic field still affects the electron acceleration dramatically. We derive an approximate analytical solution of the relativistic electron energy in adiabatic limit, which provides a full understanding of this phenomenon. Our paper is organized as follows. In Sec.II we discuss the plasma parameter used throughout the paper. In Sec.III we derive the dynamical equation describing relativistic electron in combined strong axial magnetic field and the CP electromagnetic wave field. The equation will be solved both numerically and analytically. We describe EMRA in a Gaussian CP beam with static axial magnetic field. An approximately analytical solution of relativistic electron energy is obtained, which gives a good explanation for our numerical simulation. In Sec. IV we consider the energy spectra of the accelerated particles, and the "Synchrotron radiation" will be exhibited. We summarize the results and discuss the potential applications of pulsar radio emission in Sec. V.

## II. PLASMA PARAMETERS

We choose what we consider to be the most plausible parameters. From the location of the radio emission, we are interested in the high-frequency electromagnetic wave range $\omega \approx 10^{15}$. The characteristics of pulsar circular polarization summarized by Han $et\ al.$[13] should be considered by all emission models. The polarization characteristics of the mean pulse profile provide a framework for understanding the emission processes in pulsars. The plasma rest frame density near the pulsar surface is $N_r = N_p / \gamma_p \approx 10^{11} cm^{-3}$, where $N_p \approx 10^{14} cm^{-3}$ is the resulting highly relativistic plasma density, flowing with a mean Lorentz factor of about $<\gamma> \approx 10^2$ [14]. The dipole magnetic field varies in the magnetosphere as $B \approx B_0 (R_0 / R)^{-3}$, where $B_0 = 10^{12} G$, $R_0 \approx 10^6 cm$ is the radius of the neutron star. In this paper we choose a uniform axial magnetic field in $90 MG$ for simplicity.

## III. ELECTROMAGNETIC WAVE AND STATIC-MAGNETIC FIELDS RESONANCE ACCELERATION

The approach to the analysis the response of electron in ultraintense electromagnetic wave plasma system in the presence of strong quasistatic magnetic field is in a single test model described in the relativistic Lorentz force equations

$$\frac{d\mathbf{p}}{dt} = \frac{\partial \mathbf{a}}{\partial t} - \mathbf{v} \times (\nabla \times \mathbf{a} + \mathbf{b}_z), \tag{1}$$

$$\frac{d\gamma}{dt} = \mathbf{v} \cdot \frac{\partial \mathbf{a}}{\partial t}, \tag{2}$$

where $\mathbf{a}$ is the normalized vector potential, $\mathbf{b}_z$ is the normalized static magnetic field which is parallel to the electromagnetic wave propagation direction, $\mathbf{v}$ is the normalized velocity of electron, $\mathbf{p}$ is the normalized relativistic momentum, $\gamma = (1 - v^2)^{-1/2}$ is the relativistic factor or normalized energy. Their dimensionless forms are $\mathbf{a} = \frac{e\mathbf{A}}{m_e c^2}$, $\mathbf{b} = \frac{e\mathbf{B}}{m_e c \omega}$, $\mathbf{v} = \frac{\mathbf{u}}{c}$, $\mathbf{p} = \frac{\mathbf{P}}{m_e c} = \gamma \mathbf{v}$, $t = \omega t$, $r = kr$, $m_e$ and $e$ are the electric mass and charge, respectively, $c$ is the light velocity. $k$ is the wave number. We assume that the electromagnetic wave propagation is in positive $\hat{\mathbf{z}}$ direction and moving with nearly the speed of light.

As a solution of the three-dimensional wave equation, the vector potential of an Gaussian profile electromagnetic wave can be expressed as

$$\mathbf{a} = a_0 e^{-\frac{x^2+y^2}{R_0^2}} \cdot e^{-\frac{(kz-\omega t)^2}{k^2 L^2}} \cdot [\cos(\omega t - kz)\hat{\mathbf{x}} + \delta \sin(\omega t - kz)\hat{\mathbf{y}}] \quad (3)$$
$$= a_x + \delta a_y$$

where $L$ and $R_0$ are the pulse width and minimum spot size, respectively. The two components of electromagnetic wave amplitude $\mathbf{a}$ take the form $a_x = a_0 e^{-\frac{x^2+y^2}{R_0^2}} \cdot e^{-\frac{(kz-\omega t)^2}{k^2 L^2}} \cdot [\cos(\omega t - kz)]$ and $a_y = a_0 e^{-\frac{x^2+y^2}{R_0^2}} \cdot e^{-\frac{(kz-\omega t)^2}{k^2 L^2}} \cdot [\sin(\omega t - kz)]$, respectively. $\delta$ equals to $0$, $1$, and $-1$, corresponding to linear, right-hand and left-hand circular polarization, respectively. For simplicity, we assume $\delta = 1$ (right-hand circular polarization) in the following discussions.

$\mathbf{b}_z$ is the static magnetic-field aligned to the electromagnetic wave propagation direction. We assume that the trajectory of a test electron starts at $\mathbf{r}_0 = \mathbf{v}_0 = 0$. Eqs.(1) and (2) yield

$$\frac{dp_x}{dt} = (v_z - 1)\frac{\partial a_x}{\partial z} - v_y b_z, \quad (4)$$

$$\frac{dp_y}{dt} = (v_z - 1)\delta \frac{\partial a_y}{\partial y} + v_x b_z, \quad (5)$$

$$\frac{dp_z}{dt} = -v_x \frac{\partial a_x}{\partial z} - v_y \delta \frac{\partial a_y}{\partial z}, \quad (6)$$

$$\frac{d\gamma}{dt} = -v_x \frac{\partial a_x}{\partial z} - v_y \delta \frac{\partial a_y}{\partial z}. \quad (7)$$

Using Eqs.(4)-(7), we choose different initial position to investigate the electron dynamics for a Gaussian profile electromagnetic wave pulse. Because the initial velocity can be transformed to initial position in our single test electron case, we keep initial velocity at rest and change the initial positions of the test electrons. We assume that the trajectory of a test electron starts from $\mathbf{v}_0 = 0$ and $z_0 = 4L$ at $t = 0$, while the center of electromagnetic wave pulse locates at $z = 0$, then the classical trajectory is then fully determined by Eqs.(4)-(7). Now we choose following parameters that are available in present experiments, i.e. $L = 10\lambda$, $R_0 = 5\lambda$ ($\lambda = 1.06\mu m$), $\delta = 1$, $a_0 = 4$ (corresponding $I = 2 \times 10^{19} W/cm^2$), $\mathbf{b}_z = 0.9$ (corresponding static $B_z = 90MG$), $\mathbf{r}_0 = 0.1$. Then, we trace the temporal evolution of electron energy, plot the numerical result in Fig.1.

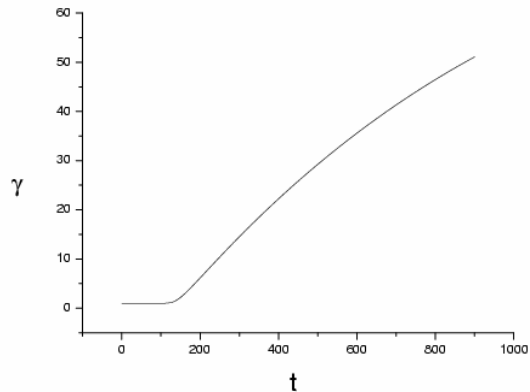

Fig.1 The electron energy $\gamma$ in units of $mc^2$ as a function of time in the units of $\omega^{-1}$ of the CP electromagnetic wave in numerical solutions by the Eqs.(4)-(7). The parameters $a_0 = 4$, $\delta = 1$, $\mathbf{b}_z = 0.9$ (corresponding to $B_z = 90MG$), $\mathbf{r}_0 = 0.01$, $\mathbf{v}_0 = 0$.

Obtaining an exactly analytical solution of Eqs.(4)-(7) is impossible because of their nonlinearity. However, we notice that the second term on the left side of Eqs.(4)-(5) possesses symmetric form, which is found to be a small quantity negligibly after careful evaluation. Then, an approximately analytical solutions of Eqs.(4)-(7) in adiabatic limit can be obtained.

From the phase of the electromagnetic wave pulse, we have the equation

$$\frac{d\eta}{dt} = \omega(1 - v_z), \quad (8)$$

where $\eta = \omega t - kz$. Then, from Eq.(10) and Eq.(11), we can easily arrive at the second useful relation under the initial condition $\mathbf{v}_0 = 0$ at $t = 0$,

$$\gamma v_z = \gamma - 1. \quad (9)$$

Finally, the energy-momentum equation yields

$$\gamma^2 = 1 + (\gamma v_x)^2 + (\gamma v_y)^2 + (\gamma v_z)^2. \quad (10)$$

The solution of the electron momentum has symmetry. In adiabatic limit and under the initial condition of zero velocity electron, we find the solutions of Eq.(4) and Eq.(5) taking the form,

$$\omega p_x = -b_z \cos(\omega t - kz), \quad (11)$$

$$\omega p_y = b_z \sin(\omega t - kz). \quad (12)$$

Substituting Eq.(11) and Eq.(12) into Eq.(4), and using Eqs.(8)-(10), we obtain an equation having a resonance point (singularity) at a positive $b_z (= \omega)$

$$\gamma = 1 + \frac{1}{2} \frac{a^2}{\left(1 - \frac{b_z}{\omega}\right)^2}. \quad (13)$$

Eq.(13) is an approximately analytical energy solutions of Eqs.(4)-(7). We show the analytical results of $\gamma \propto t$ line in Fig.2 with the same parameters and compare with the numerical result of Fig.1. The Eq.(13) can express the electron energy evolution very well.

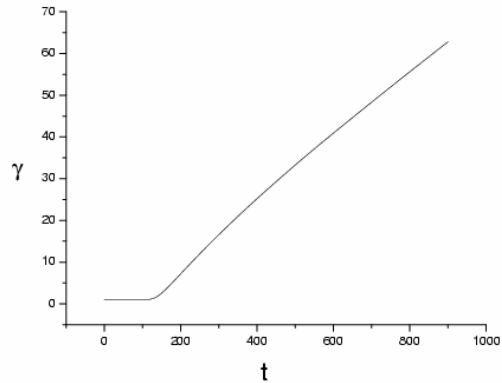

Fig.2 The $\gamma \propto t$ line of analytical formula (13) corresponding to the same parameters with Fig.1.

From the above analytic solution and our numerical calculation, we find that the strong magnetic field affect the electron acceleration dramatically through the electromagnetic and magnetic field resonance acceleration (EMRA). The electron acceleration depends not only on the electromagnetic wave intensity, but also on the ratio between electron Larmor frequency and electromagnetic wave frequency. The similar process can be happened in laser-plasma interaction system[15].

In order to get the analytical expression of electron energy $\gamma$ near the exact resonance point ($b_z = \omega$), we plug following approximate solutions into the dynamical equations,

$$\omega p_x = c(t)\sin(\omega t - kz), \quad (14)$$

$$\omega p_y = c(t)\cos(\omega t - kz), \quad (15)$$

where $c(t)$ is a coefficient to be fitted. Careful analysis gives the solution at $t \to \infty$ in the following approximate expression

$$\gamma \approx (\frac{3}{\sqrt{2}} a\omega t)^{\frac{2}{3}}. \quad (16)$$

It indicates that the resonance between the electromagnetic wave and magnetic field will drive the energy of electrons to infinity with a $2/3$ power law in time. In typical perimeter of pulsar magnetospheres, the mechanism provide chance to allow particles to increase their energies through the resonance of high magnetic field and high frequency electromagnetic wave in each electromagnetic wave period.

The dimensionless form of $b_z (= \frac{eB_z}{m_e c\omega})$ is equal to classical Larmor frequency $\Omega$ ($= \frac{eB_z}{m_e c}$). So the electron obtain energy efficiently from near or at resonance point which is the ratio of classical Larmor frequency ($\Omega = b_z$) and electromagnetic wave frequency ($\omega$).

## IV. RADIATION SPECTRA

The calculations made in Sec.III are essential for the study of radiation emitted by the accelerated electron. The starting point for calculating the frequency distribution of the radiation is the radiant energy emitted per unit frequency interval $d\omega$. The radiation from moving electrons can be analyzed in terms of the well-known Liénard-Wiechert potentials[16]. In the far-field approximation, the electron field of the radiation observed at a position $\mathbf{z}$ at time $t$ is given by

$$\mathbf{E}(\mathbf{z},t) = -\frac{e}{c^2} \frac{\mathbf{n} \times [(\mathbf{n} - \mathbf{v}/c) \times \mathbf{a}]}{(1 - \mathbf{n} \cdot \mathbf{v}/c)^3 R} \bigg|_{t=t'} \quad (20)$$

where $\mathbf{E}(\mathbf{z},t)$ is used here to denote the radiated energy, $\mathbf{n}(t)$ is a unit vector that points from the position $\mathbf{r}(t)$ of the charge towards the observer, $\mathbf{v}(t)$ the velocity, $\mathbf{a}(t)$ the acceleration of the charge and $t'$ is the retarded time defined by $c(t-t') = R(t') = |\mathbf{z} - \mathbf{r}(t')|$. For relativistic electrons the radiation is emitted into a narrow cone along the instantaneous direction of motion, where the product $\mathbf{n} \cdot v$ is close to $c$. An observer on the propagation direction therefore will receive a sequence of short radiation pulses emitted from the EMRA motion. From the Fourier transform, one obtains a broadband radiation frequency spectrum as a consequence.

Figs.3 and 4 give the spectra of power calculated numerically from Eqs.(4)-(7), corresponding to presence and absence the high magnetic field respectively. From the difference of the two figures, one can find that the high magnetic field effect the electron acceleration dramatically and make the energy spectrum exhibit the synchrotron radiation behavior.

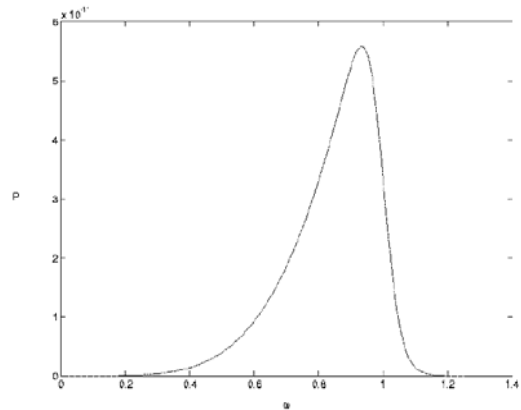

Fig.3 Displays normalized spectra of power $P$ along with the normalized frequency $\omega$ corresponding to the same parameters in Fig.1. with the high magnetic field.

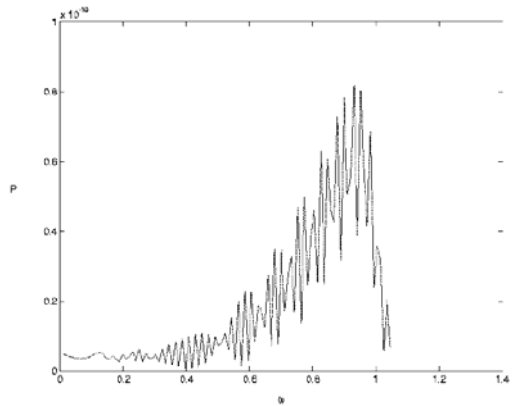

Fig.4 Displays normalized spectra of power $P$ along with the normalized frequency $\omega$ corresponding to the same parameters in Fig.1 absence the high magnetic field.

It is interesting to note that the synchrotron radiation has special characteristics not related with the electron acceleration mechanism. It produces a highly collimated, polarized, continuous spectrum, which includes wavelengths not available from other sources. This proved a key to some major current mysteries of the universe.

Detailed calculation from the radiation of EMRA mechanism show that the intensity of radiation as a function of frequency is not entirely flat over the whole range, but that it rises slowly from low values to a maximum value reached at approximately $\omega = \omega_{max}$.

## V. DISCUSSIONS AND CONCLUSIONS

We have derived relativistic energy equation for a single electron in the presence of the combined effect of magnetic field and electromagnetic wave and show its radiation spectrum, without any restrictions on the strength of the magnetic field, the intensity of the electromagnetic wave, or the initial direction of motion of the electron. The parameters can available in high energy density (HED) laser facilities which are the important aspect of the ICF-Astrophysics interaction. Using a single test electron model, we investigate the acceleration mechanism of energetic electrons in combined strong axial magnetic field and circular polarized electromagnetic wave field. An analytic solution of electron energy is obtained. We find that the electron acceleration depends not only on the electromagnetic wave intensity, known as the pondermotive acceleration, but also on the ratio between electron Larmor frequency and the electromagnetic wave frequency. As the ratio equals to unity, a clear resonance peak is observed, that is the electromagnetic and magnetic field resonance acceleration (EMRA). The strong magnetic field affects electron acceleration dramatically. This work can help to understand the continued emission of high energy electron from radio pulsars within supernova remnants.

The synchrotron radiation is being used more and more as a tool in a number of disciplines including spectroscopy, photochemistry, material studies and biology. The discovery of a pulsar within the Crab Nebula, generally thought to be the remnant of a historical supernova in 1054 AD, qualitatively solved the mystery of why continuum optical radiation from that nebula is highly polarized and what powers it. Evidently a magnetized plasma flows from the pulsar. We can conceive that the central part of the Crab nebula is still emitting high speed particles related with the pulsar radiation process. When electrons pass through matter, they suffer magnetic fields, they emit electromagnetic radiation. We must conclude that a charge moving at constant velocity cannot radiate energy. The electron moves in circular orbits suffer magnetic fields in betatron and synchrotron emissions. As a radiation source, EMRA provides a wide range and high value of research done with synchrotron radiation.

## VI. ACKNOWLEDGMENTS

HL thanks B..Qiao for useful discussion. This work was supported by National Hi-Tech Inertial Confinement Fusion Committee of China, National Natural Science Foundation of China, National Basic Research Project nonlinear Science in China, and National Key Basic Research Special Foundation.